\documentclass{amsart}
\usepackage{amssymb,amsthm,amsmath}
\usepackage{geometry}

\newcommand{\RR}{\mathbb{R}}
\newcommand{\ZZ}{\mathbb{Z}}
\newcommand{\NN}{\mathbb{N}}
\newcommand{\PP}{\mathbb{P}}
\newcommand{\EE}{\mathbb{E}}

\newcommand{\<}{\langle}

\renewcommand{\>}{\rangle}

\newtheorem{thm}{Theorem}
\newtheorem{lem}[thm]{Lemma}

\newtheorem{cor}[thm]{Corollary}

\theoremstyle{definition}
\newtheorem{dfn}[thm]{Definition}
\title{Minkowski space is locally the Noldus limit of Poisson process Causets}
\author{Jan Cristina}
\date{}
\begin{document}
\maketitle

\begin{abstract}
  A poisson process $P_{\lambda}$ on $\RR^{d}$ with causal structure inherited from the  the usual Minkowski metric on $\RR^{d}$ has a normalised discrete causal distance $D_{\lambda}(x,y)$ given by the height of the longest causal chain normalised by $\lambda^{1/d}c_{d}$.  We prove that $P_{\lambda}$ restricted to a compact set $Q$ converges in probability in the sense of Noldus \cite{noldusdistance} to $Q$ with the Minkowksi metric.
\end{abstract}

\section{Introduction}
The notion of a causal set or causet introduced in \cite{causet_start} is a proposed framework for developing a theory of quantum gravity.  It posits that spacetime can is in some sense a continuous approximation of a set of points with a partial order given by causal precedence.  It is ontologically parsimonious, and hence elegant, but being only in its nascency, it still lacks much to be a full theory of quantum gravity.  Much work has been done by examining the behaviour of random causets in Minkowski space as an attempt to derive an appropriate notion of convergence \cite{continuum_top_causet,continuum_causet} and dynamics \cite{causet_seq_dynamics}, including the promising transitive percolation model.  

Despite much work a satisfying proof of convergence of random causets has remained elusive.  For a geometric analyst used to studying convergence of metric spaces, the obvious answer is to generalise the notion of Gromov--Hausdorff convergence to causal-metric spaces.  This was done by Noldus in \cite{noldusdistance}, who constructed a distance between compact subsets of causal spaces $(Q_{1},d_{1})$ $(Q_{2},d_{2})$ presently denoted
\[d_{N}(Q_{1},Q_{2}),\]
which Noldus showed forms a complete metric on the class of compact causal spaces \cite{nolduscomplete}.

In this paper we study the convergence in the Noldus metric of a Poisson process $P_{\lambda}$ on $\RR^{d}$ with the Lorentzian metric $(dx^{1})^{2}-\sum_{i=2}^{d}(dx^{i})^{2}$. The process $P_{\lambda}$ inherits a causal structure from $\RR^{d}$, which allows us to define a discrete causal-metric on $P_{\lambda}$ given by
\[D_{\lambda}(x,y)=H\<x,y\>/(c_{d}\lambda^{d})\]
for any two points $x,y\in P_{\lambda}$, where $c_{d}$ is a number which only depends on the dimension.  We prove the following theorem:

\begin{thm}For every dimension $d\in \NN$ there are numbers $K_{d},C_{1,d}$, $C_{2,d}$ such that for every interval $Q=\<x,y\>\subset \RR^{d}$, every $\varepsilon>0$ if a density $\lambda$ satisfies
  \[K_{d}\lambda^{-1/2d}\leq \varepsilon\leq 8K_{d}\log \lambda,\]
  then
  \label{thm:convergence_in_probability}
  \[\PP(d_{N}( (P_{\lambda}\cap Q,D_{\lambda}),(Q,d))\geq \varepsilon))\leq C_{1,d}h(Q)^{2d+1}\varepsilon^{2(1-d)}\lambda^{1/d}\exp(-C_{2,d}\varepsilon^{2}\lambda^{1/d}/\log^{3}\lambda\varepsilon^{d}).\]
In particular $P_{\lambda}\cap Q\to_{N}Q$ in probability as $\lambda\to \infty$.
\end{thm}

The proof of the theorem is straightforward, and can be seen by bounding the Noldus distance to a uniform lattice, and applying the following corollary of two later theorems which is claimed by Bollob\'as and Brightwell in \cite{concentration_of_measure}: 

\begin{cor}Let $H_{\lambda}$ denote the longest height of a chain in $P_{\lambda}\cap \<0,1\>\subset \RR^{d}$. For every $d\in \NN$ there are numbers $c_{d}$ and $K_{d}$ such that for every $2\leq \mu\leq \lambda^{1/2d}/\log\log \lambda$.
  \label{cor:concentration_of_measure}
  \[\PP\left(|H_{\lambda}-\lambda^{1/d}c_{d}|\geq \frac{K_{d}\lambda^{1/(2d)}\log^{3/2}\lambda}{\log \log \lambda}\right)\leq 4\mu^{2}\exp(-\mu^{2})\]
\end{cor}
\proof This follows directly by combinging Theorems \ref{thm:deviation_from_mean} and \ref{thm:convergence_of_mean} noting that $\EE H_{\lambda}= c_{\lambda} \lambda^{1/d}$. 
Theorems \ref{thm:deviation_from_mean} and \ref{thm:convergence_of_mean} are direct generalisations of the corresponding result of Bollob\'as and Brightwell, which as they state only requires a little effort (provided here for the benefit of the reader).\endproof

The idea of using the height of maximal chains to define a timelike distance, is essentially quantised proper time, i.e. proper time is allowed to take values only in $\eta\NN$, for some scale factor $\eta$.  
\section{Preliminaries}
We consider consider Minkowski space $\RR^{d}$ with the Lorentzian metric
\[ds^{2}=(dx^{1})^{2}-\sum_{i=2}^{d}(dx^{i})^{2}.\]
This defines a natural causal-metric on $\RR^{d}$ by
\[d(x,y)=\sqrt{\max\left\{0,(y^{1}-x^{1})^{2}-\sum_{i=2}^{d}(y^{i}-x^{i})^{2}\right\}}\]
\begin{dfn}For our purposes a causal distance $d:X\times X\to \RR^{+}$ satisfies the following three properties
\begin{enumerate}
  \item For every $x\in X$ \[d(x,x)=0,\]
  \item For every $x,y\in X$ if $d(x,y)>0$ then $d(y,x)=0$, and
  \item if $d(x,y)>0$ and $d(y,z)>0$ then
\[d(x,y)+d(y,z)\leq d(x,z).\]
\end{enumerate}
\end{dfn}
Although causal distances superficially resemble metrics, the reverse triangle inequality makes working with them highly nonintuitive.

We can define the timelike height of a set $Q$ to be 
\[h(Q)=\sup_{x,y\in Q}d(x,y).\]
Most of our results will be contingent on the height of the set under investigation.  The notion will also be useful in rescaling the sets under consideration to a standard reference set.

We define the causal future of a point $x$ to by
\[J_{+}(x)=\{y:d(x,y)>0\},\]
and the causal past of $y$ to be 
\[J_{-}(y)=\{x:d(x,y)>0\}.\]
Then for any two points $x,y$ we define the spacetime interval 
\[\<x,y\>=J_{+}(x)\cap J_{-}(y).\]
In particular for Minkowski space 
\[h(\<x,y\>)=d(x,y).\]

Because Minkowksi space is preserved under Lorentz transformations, it is useful to introduce the Minkowski diamond, which is the set
\[\<0,t\>=\<0,(t,0)\>.\]
This slight abuse of notation is introduced for legibility's sake, as the latter while more technically correct is less readable and less intuitive than the former.

Most importantly the standard Minkowski diamond is the set $\<0,1\>$.  This has $d$-dimensional Lebesgue measure $C_{d}$, i.e.
\[|\<0,1\>|=C_{d}.\]
Consequently, because every non-empty spacetime interval $\<x,y\>$ can be translated and then Lorentz boosted into a Minkowski diamond of the form $\<0,d(x,y)\>$, it follows that
\[|\<x,y\>|=C_{d}d(x,y)^{d}.\]

Given two causal metric spaces $(X_{1},d_{1})$ and $(X_{2},d_{2})$ we say they are $\varepsilon$ close if there are maps $\psi:X_{1}\to X_{2}$ and $\varphi:X_{2}\to X_{1}$ such that
\[\sup_{x,y\in X_{1}}|d_{1}(x,y)-d_{2}(\psi(x),\psi(y))|\leq \varepsilon\]
and
\[\sup_{x,y\in X_{2}}|d_{2}(x,y)-d_{1}(\varphi(x),\varphi(y))|\leq \varepsilon.\]
The Noldus distance is
\[d_{N}( (X_{1},d_{1}),(X_{2},d_{2}))=\inf\{\varepsilon:(X_{1},d_{1})\text{ and }(X_{2},d_{2})\text{ are $\varepsilon$ close}\}.\]
For brevity's sake we will omit the causal distance functions when it is clear, i.e.
\[d(X_{1},X_{2})\text{ will mean }d((X_{1},d_{1}),(X_{2},d_{2}))\]
when $X_{1}$ and $X_{2}$ can be obviously (within the context of this article) be assigned causal metrics.

Noldus introduced his notion of distance \cite{noldusdistance}, as a generalisation of Gromov's notion of Gromov--Hausdorff distance for metric spaces.  The important properties proved in \cite{noldusdistance} and \cite{nolduscomplete} are that this is indeed a metric on the class of compact causal metric spaces.

\section{Causets and Poisson Processes}
A causal set or causet, is a set $X$ with a partial order $\leq$.  In other words it is a partially ordered set or poset.  We say that $x<y$ if $x\leq y$ and $x\neq y$.  A causet has a natural causal distance given by 
\[D(x,y)=\sup_{N}x<x_{1}<\ldots<x_{N}<y.\]
Causets were proposed as a potential perspective for quantum gravity in \cite{causet_start}.  Although causets are posets, it seems beneficial to refer to them as causets in the context of quantum gravity, to emphasise their relation to causality.

One idea of the causet program has been to use Poisson processes as a toy model of discretised spacetime.  The primary benefit of a Poisson process, as opposed to deterministic discretisation of spacetime, is that it is Lorentz invariant, i.e. given a Lorentz boost $T$, and a Poisson process with density $\lambda$, $P_{\lambda}$, $T(P_{\lambda})$ is also a Poisson process with density $\lambda$.

A Poisson process with density $\lambda$ is defined by the following properties, 
\begin{enumerate}
  \item The probability of finding $n$ points in a set $A$ is given by
    \[\PP(|A\cap P_{\lambda}|=n)=e^{-\lambda|A|}(\lambda|A|)^{n}/n!,\]
    where $|\cdot|$ is either the cardinality or Lebesgue measure appropriately.

  \item For $A$ and $B$ disjoint sets $A\cap P_{\lambda}$ $B\cap P_{\lambda}$ are independent random variables.
\end{enumerate}

A poisson process can be constructed by subdividing a $\sigma$-finite measure space $X,\mu$ into countable many disjoint sets of finite measure $A_{i}$ and for each $A_{i}$ defining a Poisson randomvariable $N_{i}$ distributed like $\lambda|A_{i}|$, and $X_{1}^{i},\ldots ,X^{i}_{N_{i}}$ uniform random variables in $A_{i}$

Given a Poisson process with density $\lambda$, $P_{\lambda}$ on a causal metric space $(X,d)$ equipped with a $\sigma$-finite measure $\mu$, $P_{\lambda}$ inherits a causal structure from $(X,d)$ via $x\leq y$ if and only if $x=y$ or $d(x,y)>0$.  Poisson processes were used as tools to study random partial orders.  The field is vigorous, and active, but the relevant citations for this paper are Bollob\'as and Brightwell papers \cite{concentration_of_measure,box_spaces}

For a finite causet, we can define the height $H(Q)$ of a subset $Q$ to be 
\[H(Q)=\max\{N:\:\exists\: x_{1}<\ldots<x_{N}\in Q\}. \]
For the standard Minkowski diamond, we introduce a random variable $H_{\lambda}=H(P_{\lambda}\cap \<0,1\>)$.   An important property that follows immediately from Lorentz invariance and the scaling properties of Poisson processes us
\[H(P_{\lambda}\cap\<x,y\>)\sim H_{\lambda d(x,y)^{d}}.\]

\section{Concentration of measure}

The goal of this section is to  apply a ``little effort'' and  modify the proof of Bollob\'as and Brightwell in \cite{concentration_of_measure} to apply to Minkowski space as they claim following their Theorem 1.  The first step is to modify the proof of Theorem 3 from the case of the Cartesian order to that of the standard Minkowski diamond. The only modifications are minor modifications in certain quantities.

\begin{thm}Let $H_{\lambda}=H(P_{\lambda}\cap \<0,1\>)$.  For every $2\leq \mu\leq \lambda^{1/2d}\log \lambda$
  \label{thm:deviation_from_mean}
  \[\PP \left( |H_{\lambda}-\EE H_{\lambda}|> \frac{\mu K_{d}\lambda^{1/(2d)}\log \lambda}{\log \log \lambda} \right)\leq 4\mu^{2}\exp(-\mu^{2})\]
\end{thm}
\proof
One difference between the Minkowksi diamond and the Cartesian order is the division of strips.  In Minkowksi space we just divide the time coordinate $x_{0}$ instead of the sum $\sum_{i}x_{i}$. The strips are defined as $X_{j}=\{(x_{0},\ldots,x_{d-1}):\leq (j-1)/m x_{0}\leq j/m\}$, and applying Lemma 4 from \cite{concentration_of_measure}.  

Now following Bollob\'as and Brightwell we define a new random variable \[H'=\text{ the length of the longest chain such $C$ such that $|C\cap S_{j}|\leq 2^{d+1}\log \lambda/\log \log \lambda$}.\]
Note in the case of the Cartesian order there is a $2(d+1)$ as opposed to $2^{d+1}$.  We let $k=2^{d+1}\log\lambda/\log \log \lambda$.  This follows because the positive light cone from a point intersected with a strip $X_{j}$ intersects at most $2^{d}$ cubes of side length $1/m$. The proof of Lemma 7 in \cite{concentration_of_measure} is unchanged for Minkowski space, because it is just a claim on the size of deviation for a Poisson Process over a subset of cubes.

\endproof
Now we can prove prove the convergence of the expected height.  Let
\[c_{\lambda}:=\frac{\EE H_{\lambda}}{n^{1/d}}.\]  From \cite{box_spaces} we know $c_{\lambda}\to c$.
\begin{thm}
  \label{thm:convergence_of_mean}
  For $\lambda$ sufficiently large
  \[c\geq c_{\lambda}\geq c_{\lambda}-\frac{C_{d} \log^{3/2}\lambda}{\lambda^{1/2d}\log\log \lambda}.\]
\end{thm}
\proof  As in \cite{concentration_of_measure}, we can split the interval$\<0,2\>$ into $\<0,1\>$ $\<1,2\>$, yielding
\[\EE(H_{2^{d}\lambda})\geq 2 \EE(H_{\lambda}).\]
Then as before we consider the longest chain $C$, and let $x$ denote it's midpoint.  Let $h$ be the minimum of $d(0,x)$ and  $d(x,2)$.  Consequently $h\leq 1$.  Without loss of generality assume that $h=d(0,x)$, hence $|\<0,x\>|\leq |\<0,1\>|$  and hence $H_{\lambda}(\<0,x\>)\leq H_{\lambda}(\<0,1\>)$.  Thus 
\begin{align*}
  \PP(x&\text{ is a midpoint of a chain of length }\geq 2 \EE_{\lambda,d}+2\mu\lambda^{1/2d}\log \lambda/\log\log \lambda)\\
  &\leq\PP(H(\<0,x\>)\geq H_{\lambda}+\mu K_{d}\lambda^{1/2d}\log \lambda/\log\log \lambda) \\
  &\qquad\qquad\PP(H(\<x,2\>)\geq H_{\lambda}+\mu K_{d}\lambda^{1/2d}\log \lambda/\log\log \lambda)\\
  &\leq \PP(|H_{\lambda h^{d}}(\<0,1\>)-\EE H_{\lambda h^{d}}|\geq \mu K_{d}\lambda^{1/2d} h^{1/2}\log (h^{d}\lambda)/\log\log (h^{d}\lambda))\\
  &\leq 4\mu^{2}\exp(-\mu^{2}).
\end{align*}
The rest of the proof proceeds as in \cite{concentration_of_measure}.
\endproof
\section{Convergence in probability}

Let $P_{\lambda}$ be a Poisson process of density $\lambda$ on $\RR^{d}$.  Let $Q_{\lambda}$ denote $P_{\lambda}\cap Q$ for every subset $Q\subset\RR^{d}$.  We define a causal distance on $\RR^{d}$ by 
\[D_{\lambda}(x,y):=H(\<x,y\>\cap Q_{\lambda})/(c_{d}\lambda^{1/d}).\]
Restricted to $Q_{\lambda}$ this satisfies the required properties.  Let $h=h(Q)$

Let $\eta=\varepsilon^{2}/16\sqrt{d}h$, let $B_{\eta}(Q)=\{x\in \RR^d:\exists y\in Q\, |x-y|<\eta\}$, and let $\Lambda_{\varepsilon}=\eta\ZZ\cap B_{\eta}(Q)$.
\begin{lem}For every $\varepsilon>0$, the lattice $\Lambda_{\varepsilon}$ satisfies
  \label{lem:lattic_props}
  \begin{enumerate}
    \item for every $x\in Q$ there are points $x^{+}$ and $x_{-}$ satisfying $d(x_{-},x_{+})\leq 2\sqrt{d}\eta$, $|x_{+}-x_{-}|\leq 2\sqrt{d}\eta$, and  $x\in \<x_{-},x_{+}\>$;
    \item  For every pair of points $x,y\in Q$ we have
      \[d(x_{-},y_{+})\leq d(x_{+},y_{-})+\varepsilon/4;\]
    \item For every pair of points $x,y\in Q$ we have
      \[d(x_{+},y_{-})\leq d(x,y)\leq d(x_{-},y_{+});\]
    \item For every $x,y\in \Lambda$ either $d(x,y)=0$ or $d(x,y)\geq \varepsilon^{2}/4\sqrt{d}h$.
\end{enumerate}
\end{lem}
\proof
\begin{enumerate}
  \item
Given a point $x\in Q$ choose the euclidean nearest point in $\Lambda$, $x_{\eta}=(x^{0}_{\eta},\tilde{x}_{\eta})$.  Then $x^{+}=x^{0}_{\eta}+\sqrt{d}\eta,\tilde{x}_{\eta}$ and $x_{-}=x^{0}_{\eta}-\sqrt{d} \eta, \tilde{x}_{\eta}$ are candidates satisfying the condition.
\item Here we consider the maximum of $\sqrt{n_{1}^{2}-\xi^{2}}-\sqrt{n_{2}^{2}-\xi^{2}}$,  where $n_{i}\in \ZZ$ and $\xi_{i}\in \ZZ^{d-1}$.
subject to $n^{2}_{i}> \xi^{2}$.  Assume further that $n_{1}=n_{2}+4\lceil\sqrt{d}\rceil.$  Then the difference is
\[\sqrt{4n+16+h}-\sqrt{h}.\]
Thus the maximal difference is
\[\sqrt{4n+16}.\]
Now if we scale by $\eta$, and note that the maximum for $n+4$ is $(h(Q)+4\eta)/\eta$, we get the result.
\item This follows from $x_{-}\leq x\leq x_{+}$ and $y_{-}\leq y\leq y_{+}$.
\item This is because the the lattice the distance will always be given by $\eta\sqrt{z}$ for $z\in \NN$.
  \end{enumerate}
\endproof

We now bound the noldus distance of our lattice with the random causal metric, compared to 
\begin{lem}
  For every $d$ there are numbers $C_{1,d}$ and $C_{2,d}$ such that for every $\varepsilon>0$ and $\lambda$  satisfying $K_{d}\lambda^{-1/2d}\leq \varepsilon\leq 8K_{d}\log \lambda $
\label{lem:lattice_dist}
\[
  \PP(\sup_{x,y\in \Lambda}|d(x,y)-D_{\lambda}(x,y)|\geq\varepsilon/2)\leq  C_{1,d}h(Q)^{2d+1}\varepsilon^{-2(d-1)} \lambda^{1/d} e^{-C_{2,d}\varepsilon^{2}\lambda^{1/d}/\log^{3}\lambda \varepsilon^{d}}
\]
\end{lem}
\proof
We merely coarsely estimate, noting that $d(x,y)\geq \epsilon^{2}/4\sqrt{d}$ and hence $|\<x,y\>|\geq \epsilon^{2d}/4^{d}d$, so for any pair of points $x,y$ $d(x,y)\geq 0$, it follows that 
\begin{multline*}
\PP(|H\<x,y\>/\lambda^{1/d}c_{d}-d(x,y)|\geq \varepsilon/4)\leq\\ \PP\left( |H_{\lambda d(x,y)^{d}}/c_{d}d(x,y)\lambda^{1/d}-1|\geq \mu K_{d}\frac{\log^{3/2}\lambda d(x,y)^{d}}{\lambda^{1/2d}\sqrt{d(x,y)}\log\log \lambda d(x,y)^{d}} \right),
\end{multline*}
where $\mu=\varepsilon\lambda^{1/2d}\sqrt{d(x,y)}\log \log \lambda d(x,y)^{d}/8 K_{d}\log^{3/2}\lambda d(x,y)^{d}$.  Consequently for $\lambda$ sufficiently large this is bounded by 
\[\mu^{2}e^{-\mu^{2}}\leq \varepsilon^{2}\lambda^{1/d}h(Q) K_{d}^{-2} e^{-\lambda^{1/d}\epsilon^{2}/\log^{3/2}\lambda\varepsilon^{2d}} \]
We coarsely estimate the number of pairs of points as proportional to $h(Q)^{d}/\eta^{d}=4^{2d}d^{d} h^{2d}/\varepsilon^{2d}$ to yield the result. 
\endproof

  We are now equipped to 
  \proof[prove Theorem \ref{thm:convergence_in_probability}] We consider the inclusion map $Q_{\lambda}\to Q$ and the map $Q\to Q_{\lambda}$, which takes $x$ to the Euclidean nearest point in $Q_{\lambda}\cap \<x_{-},x_{+}\>$. We note that
  \begin{align*}
    D_{\lambda}(x_{+},y_{-})\leq &D_{\lambda}(x,y) \leq D_{\lambda}(x_{-},y_{+})|
    \end{align*}
    and hence
  \begin{align*}
    d(x,y)-&|d(x,y)-d(x_{+},y_{-})|-|D_{\lambda}(x_{+},y_{-})-d(x_{+},y_{-})|\\
    &\leq D_{\lambda}(x,y)\leq d(x,y)+|d(x_{-},y_{+})-d(x,y)|+|d(x_{-},y_{+})-D_{\lambda}(x_{-},y_{+})|.
    \end{align*}
    Thus the event $\sup_{x,y\in Q_{\lambda}}|d(x,y)-D_{\lambda}(x,y)|\geq \varepsilon$ is implied by the event $\sup_{x,y\in Q_{\lambda}}|d(x_{+},y_{-})-D_{\lambda}(x_{+},y_{-})|\leq \varepsilon/4,$ because of Lemma \ref{lem:lattic_props} and hence
  \begin{equation}
    \PP(\sup_{x,y\in Q_{\lambda}}|d(x,y)-D_{\lambda}(x,y)|\geq \varepsilon)\leq C_{1,d}h(Q)^{2d+1}\varepsilon^{-2(d-1)}\lambda^{1/d}e^{C_{2,d}\varepsilon^{2}\lambda^{1/d}/\log\lambda^{3}\varepsilon^{2d}} 
    \label{eqn:Noldus_dist_1}
  \end{equation}
  
Note that
\begin{align*}
  |d(x,y)-D_{\lambda}(\psi(x),\psi(y))|&\leq |d(x,y)-d(\psi(x),\psi(y))|+|d(\psi(x),\psi(y)-D_{\lambda}(\psi(x),\psi(y))|\\
  &\leq \varepsilon/2+|d(\psi(x),\psi(y))-D_{\lambda}(\psi(x),\psi(y))|.
\end{align*}
Consequently the event $\sup_{x,y\in Q}|d(x,y)-D_{\lambda}(\psi(x),\psi(y))|\geq \varepsilon$ is implied by 
\[\sup_{x,y\in Q_{\lambda}}|d(x,y)-D_{\lambda}(x,y)|\leq \varepsilon/2.\]
The proof then follows by applying  \eqref{eqn:Noldus_dist_1} with $\varepsilon$ replaced by $\varepsilon/2$.
\endproof

The proof of Theorem \ref{thm:deviation_from_mean} should generalise rather naturally to the case of a Poisson process to a Lorentzian manifold.  One could use the time coordinate for a stationary spacelike slice, and subdivide that.  Similarly the division into cubes could be done under sufficiently regular coordinates, to guarantee the appropriate bounds on the probability of deviation.    

Theorem \ref{thm:convergence_of_mean} will prove more challenging to generalise.  The self similarity of a Poisson process to itself on smaller subsets was integral in the proof.  This will not hold in general for a Lorentzian manifold.  Nonetheless by virtue of being Lorentzian, on sufficiently small scales everything would be well approximated by the Minokowski case.  One avenue of attack would thus be to try to cover a timelike curve in a Lorentzian manifold by a series of spacetime intervals.  As the size of the intervals gets smaller they will be better approximated by the Mikowskian case, and this should give a bound on the probability of deviation of a maximal chain in the causet, from the geodesic in the underlying manifold.

The notion of Noldus convergence clearly formalises an intuitive notion of convergence of causets, allowing several questions to be amenable to mathematical analysis, but in order to extrapolate appropriate dynamics for causets one would need to ascribe an appropriate notion of convergence for certain other differential geometric, and operator theoretic concepts, like curvature, spinor fields, connections, the d'Alembertian, etc. \cite{discrete_dAlembertian}.  Some interesting work in the Riemannian case could pave the way \cite{discrete_DG}.  Therein a Gromov--Hausdorff convergence is combined with a notion of convergence of the Tangent space, to derive convergence of geodesics, curvature, and the Laplace operator.  Although the framework is not immediately generalisable, it provides strong hints of an appropriate direction.

\bibliographystyle{plain}

\begin{thebibliography}{10}

\bibitem{discrete_dAlembertian}
Alessio Belenchia, Dionigi M.~T. Benincase, and Fay Dowker.
\newblock The continuum limit of a 4-dimensional causal set scalar
  d'alembertian.
\newblock {\em arXiv}, 510.04656v1, 2015.

\bibitem{box_spaces}
B{\'e}la Bollob{\'a}s and Graham Brightwell.
\newblock Box-spaces and random partial orders.
\newblock {\em Trans. Amer. Math. Soc.}, 324(1):59--72, 1991.

\bibitem{concentration_of_measure}
B{\'e}la Bollob{\'a}s and Graham Brightwell.
\newblock The height of a random partial order: concentration of measure.
\newblock {\em Ann. Appl. Probab.}, 2(4):1009--1018, 1992.

\bibitem{causet_start}
Luca Bombelli, Joohan Lee, David Meyer, and Rafael~D. Sorkin.
\newblock Space-time as a causal set.
\newblock {\em Phys. Rev. Lett.}, 59:521--524, Aug 1987.

\bibitem{discrete_DG}
Klaus Hildebrandt, Konrad Polthier, and Max Wardetzky.
\newblock On the convergence of metric and geometric properties of polyhedral
  surfaces.
\newblock {\em Geom. Dedicata}, 123:89--112, 2006.

\bibitem{continuum_top_causet}
Seth Major, David Rideout, and Sumati Surya.
\newblock On recovering continuum topology from a causal set.
\newblock {\em Journal of Mathematical Physics}, 48(3), 2007.

\bibitem{nolduscomplete}
Johan Noldus.
\newblock The limit space of a {C}auchy sequence of globally hyperbolic
  spacetimes.
\newblock {\em Classical Quantum Gravity}, 21(4):851--874, 2004.

\bibitem{noldusdistance}
Johan Noldus.
\newblock A {L}orentzian {G}romov-{H}ausdorff notion of distance.
\newblock {\em Classical Quantum Gravity}, 21(4):839--850, 2004.

\bibitem{causet_seq_dynamics}
D.~P. Rideout and R.~D. Sorkin.
\newblock Classical sequential growth dynamics for causal sets.
\newblock {\em Phys. Rev. D (3)}, 61(2):024002, 16, 2000.

\bibitem{continuum_causet}
David Rideout and Petros Wallden.
\newblock Emergent continuum spacetime from a random, discrete, partial order.
\newblock {\em Journal of Physics: Conference Series}, 189(1):012045, 2009.

\end{thebibliography}
\def\cprime{$'$}

\end{document}